# Riemann Hypothesis as an Uncertainty Relation

R. V. Ramos

Department of Teleinformatic Engineering, Federal University of Ceara
Campus do Pici, C. P. 6007, 60455-740, Fortaleza, Brazil.

rubens.viana@pq.cnpq. br

**Abstract.** Physics is a fertile environment for trying to solve some number theory problems. In particular, several tentative of linking the zeros of the Riemann-zeta function with physical phenomena were reported. In this work, the Riemann operator is introduced and used to transform the Riemann's hypothesis in a Heisenberg-type uncertainty relation, offering a new way for studying the zeros of Riemann's function.

## 1. Introduction

Riemann's hypothesis states that all the non-trivial zeros of the zeta function $\zeta(s) = \Sigma_n(1/n^s)$ are of the form $s = ½ + it$ (the trivial zeros are the negative even integers). From the pure mathematical point of view, to prove this hypothesis is a very challenging task. On the other hand, several physical systems are related to the zeros of the Riemann-zeta function [1-3], what led scientists to look for a physical reason that forbids the existence of zeros that does not lie in the critical line, Re($s$) = ½ . Here, I introduce the Riemann operator aiming to transform the Riemann hypothesis in a Heisenberg-type uncertainty relation. It is shown that the presence of non-trivial zeros out of the critical line may lead to violation of that uncertainty relation. Firstly, let me assume that the Riemann-zeta function can be written as a series

$$\zeta(s) = \sum_{k=0}^{\infty} C_k s^k. \qquad (1)$$

Changing the complex variable $s$ by the annihilation operator $a$, one gets the Riemann operator $\zeta = \Sigma_k C_k a^k$. Since $a|\alpha\rangle = \alpha|\alpha\rangle$ where $|\alpha\rangle$ is a coherent state, one has $\zeta|\alpha\rangle = \zeta(\alpha)|\alpha\rangle$. Before using the Riemann operator, it is useful to remind the following issues: 1) $D(\alpha)|\beta\rangle = \exp(i\mathrm{Im}(\alpha\beta^*))|\alpha+\beta\rangle$ where $D(\bullet)$ is the Glauber's displacement operator. 2) the Heisenberg uncertainty relation $(\Delta A)^2(\Delta B)^2 \geq 0.25|\langle[A,B]\rangle|^2$. This work uses coherent states, hence, $(\Delta A)^2 = \langle\alpha|A^2|\alpha\rangle - \langle\alpha|A|\alpha\rangle^2$, $(\Delta B)^2 = \langle\alpha|B^2|\alpha\rangle - \langle\alpha|B|\alpha\rangle^2$ and $\langle[A,B]\rangle = \langle\alpha|AB-BA|\alpha\rangle$.

## 2. Riemman hypothesis as an uncertainty relation

Now, consider the following Hermitean operators,

$$X_1(\beta) = D^\dagger(\beta)(\zeta + \zeta^\dagger)D(\beta) \tag{2}$$

$$X_2 = -i(\zeta - \zeta^\dagger). \tag{3}$$

The uncertainty relation for the Hermitean operators $X_1(\beta)$ and $X_2$ calculated in the state $|\alpha\rangle$ is,

$$(\Delta X_1(\beta))^2 (\Delta X_2)^2 = \left(\langle\alpha+\beta|\zeta\zeta^\dagger|\alpha+\beta\rangle - |\zeta(\alpha+\beta)|^2\right)\left(\langle\alpha|\zeta\zeta^\dagger|\alpha\rangle - |\zeta(\alpha)|^2\right) \geq \\
\text{Re}^2\left[\zeta(\alpha+\beta)\zeta^*(\alpha) - e^{-i\text{Im}(\alpha\beta^*)}\langle\alpha+\beta|\zeta D(\beta)\zeta^\dagger|\alpha\rangle\right]. \tag{4}$$

The uncertainty relation in (4) must be valid for any values of $\alpha$ and $\beta$. Now, using $\alpha = (1+\varepsilon)/2 + it$ and $\beta = (1+\varepsilon)/2 - it$, where $\alpha$ and $\beta$ are supposed to be zeros of the Riemann-zeta function, $\zeta((1+\varepsilon)/2 + it) = \zeta((1+\varepsilon)/2 - it) = 0$, the uncertainty relation is

$$\left(\langle 1+\varepsilon|\zeta\zeta^\dagger|1+\varepsilon\rangle - |\zeta(1+\varepsilon)|^2\right)\langle(1+\varepsilon)/2+it|\zeta\zeta^\dagger|(1+\varepsilon)/2+it\rangle \geq \\
\text{Re}^2\left[\zeta(1+\varepsilon)\zeta^*((1+\varepsilon)/2+it) - e^{-i(1+\varepsilon)t}\langle 1+\varepsilon|\zeta D\zeta^\dagger|(1+\varepsilon)/2+it\rangle\right]. \tag{5}$$

Once zeros of the type ½ + *it* are known to exist, the inequality in (5) must be obeyed when $\varepsilon = 0$ and *t* assuming the correct values. Now, let me assume there exist zeros out of the critical line. For a zero with $\varepsilon > 0$, the term $\zeta(1+\varepsilon)\zeta^*((1+\varepsilon)/2 + it)$ in (5) vanishes and the uncertainty relation is simplified to

$$\left(\langle 1+\varepsilon|\zeta\zeta^\dagger|1+\varepsilon\rangle - |\zeta(1+\varepsilon)|^2\right)\langle(1+\varepsilon)/2+it|\zeta\zeta^\dagger|(1+\varepsilon)/2+it\rangle \geq \text{Re}^2\left[e^{-i(1+\varepsilon)t}\langle 1+\varepsilon|\zeta D\zeta^\dagger|(1+\varepsilon)/2+it\rangle\right]. \tag{6}$$

Once the values for $\varepsilon$ and *t* were chosen, if the inequality in (6) is false then zeros with values $(1+\varepsilon)/2 \pm it$ cannot exist. On the other hand, having the inequality in (6) satisfied for a pair of values of $\varepsilon$ and *t*, does not necessarily imply that $(1+\varepsilon)/2 \pm it$ are zeros. Rewriting (6) in the form

$$\left(\langle 1+\varepsilon|\zeta\zeta^\dagger|1+\varepsilon\rangle - |\zeta(1+\varepsilon)|^2\right) \geq \frac{\text{Re}^2\left[e^{-i(1+\varepsilon)t}\langle 1+\varepsilon|\zeta D\zeta^\dagger|(1+\varepsilon)/2+it\rangle\right]}{\langle(1+\varepsilon)/2+it|\zeta\zeta^\dagger|(1+\varepsilon)/2+it\rangle} \Rightarrow f(\varepsilon) \geq g(t;\varepsilon), \tag{7}$$

one can note the left side depends only on $\varepsilon$ while the right side depends on both *t* and $\varepsilon$, the last working as a parameter. Hereafter it is assumed that $g(t;\varepsilon)$ is continuous for $\varepsilon > 0$ (what is reasonable for a real physical system) and that the complex number $(1+\varepsilon)/2 + it_1$ is a zero, hence, $f(\varepsilon) \geq g(t_1,\varepsilon)$. On the other hand, given a value for $\varepsilon > 0$, the inequality in (7) cannot be true for all values of *t*. This happens because (7) is only a special case of (4) that, by its turn, must be always true. Thus, since the inequality in (7) is not always satisfied and the function $g(t;\varepsilon)$ is continuous, there must exist two values for *t*, $t_2$ and $t_3$, such that $f(\varepsilon) = g(t_2;\varepsilon)$ and $f(\varepsilon) < g(t_3;\varepsilon)$. Varying *t*, the curve of $g(t;\varepsilon)$ must pass by those points, as shown in Fig. 1.

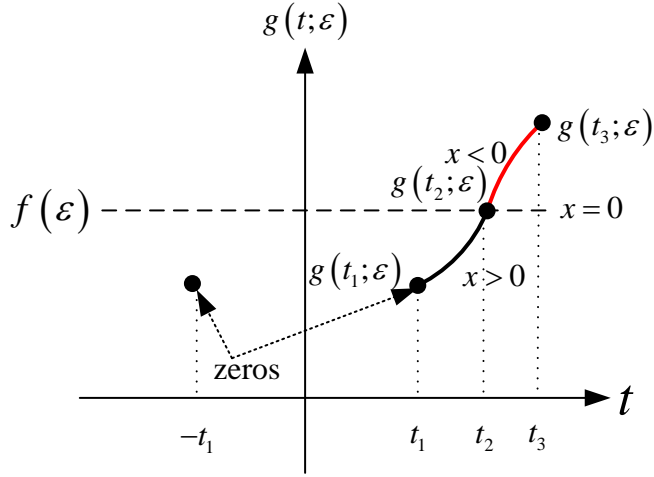

Fig. 1 – A sketch of $g(t;\varepsilon)$ versus $t$ considering the possible existence of the zero $(1+\varepsilon)/2 + it_1$ out of the critical line.

Now, it is possible to rewrite (6) in the form

$$\left(\langle 1+\varepsilon|\zeta\zeta^\dagger|1+\varepsilon\rangle - |\zeta(1+\varepsilon)|^2\right)\langle (1+\varepsilon)/2+it|\zeta\zeta^\dagger|(1+\varepsilon)/2+it\rangle - x(t) = \operatorname{Re}^2\left[e^{-i(1+\varepsilon)t}\langle 1+\varepsilon|\zeta D\zeta^\dagger|(1+\varepsilon)/2+it\rangle\right], \quad (8)$$

where $x(t)$ is a positive or null (negative) value if (6) is (not) satisfied. Consider the point $(1+\varepsilon)/2+it'$, $t_1 < t' \leq t_3$ in Fig.1. The uncertainty relation given in (4), with $\beta = (1+\varepsilon)/2 - it'$, is

$$\left(\langle 1+\varepsilon|\zeta\zeta^\dagger|1+\varepsilon\rangle - |\zeta(1+\varepsilon)|^2\right)\left[\langle (1+\varepsilon)/2+it'|\zeta\zeta^\dagger|(1+\varepsilon)/2+it'\rangle - |\zeta((1+\varepsilon)/2+it')|^2\right] \geq \operatorname{Re}^2\left[\zeta(1+\varepsilon)\zeta^*((1+\varepsilon)/2+it')\right]$$
$$+ \operatorname{Re}^2\left[e^{-i(1+\varepsilon)t'}\langle 1+\varepsilon|\zeta D\zeta^\dagger|(1+\varepsilon)/2+it'\rangle\right] - 2\operatorname{Re}\left[\zeta(1+\varepsilon)\zeta^*((1+\varepsilon)/2+it')\right]\operatorname{Re}\left[e^{-i(1+\varepsilon)t'}\langle 1+\varepsilon|\zeta D(\beta)\zeta^\dagger|(1+\varepsilon)/2+it'\rangle\right]. \quad (9)$$

Substituting (8), with $t = t'$, in (9) one gets

$$0 \geq \operatorname{Re}^2\left[\zeta(1+\varepsilon)\zeta^*((1+\varepsilon)/2+it')\right] + \left(\langle 1+\varepsilon|\zeta\zeta^\dagger|1+\varepsilon\rangle - |\zeta(1+\varepsilon)|^2\right)|\zeta((1+\varepsilon)/2+it')|^2 - x(t')$$
$$-2\operatorname{Re}\left[\zeta(1+\varepsilon)\zeta^*((1+\varepsilon)/2+it')\right]\sqrt{\left(\langle 1+\varepsilon|\zeta\zeta^\dagger|1+\varepsilon\rangle - |\zeta(1+\varepsilon)|^2\right)\langle (1+\varepsilon)/2+it|\zeta\zeta^\dagger|(1+\varepsilon)/2+it'\rangle - x(t')}. \quad (10)$$

Observing (10) one sees the first and the second terms of the right side, as well the term inside of the square root sign, are always positives. Furthermore, $\zeta(1+\varepsilon)$ is always real and positive in the range $0 < \varepsilon \leq 1$. Thus, in several situations (10) is not satisfied. In particular, depending on the value of $x(t')$, the inequality in (10) may not be satisfied if $\operatorname{Re}[\zeta^*((1+\varepsilon)/2+it')] \leq 0$. For example, for $t_2 \leq t' \leq t_3$ one has $x(t') \leq 0$ and, in this case, the inequality in (10) will not be satisfied if $\operatorname{Re}[\zeta^*((1+\varepsilon)/2+it')] \leq 0$. Hence, the existence the point $(1+\varepsilon)/2 + it'$, $t_2 \leq t' \leq t_3$ with $\operatorname{Re}[\zeta((1+\varepsilon)/2 + it')] \leq 0$, implies that the zero $(1+\varepsilon)/2 + it_1$ cannot exist because its existence implies in the existence of the coherent state $|(1+\varepsilon)/2 + it'\rangle$ for which the uncertainty relation in (4) is not valid.

On the other hand, Bohr and Courant proved [4] that for any fixed $\alpha \in (1/2,1]$ the set of values of $\zeta(\alpha+it)$ with $t \in \Re$ lies dense in the complex plane, implying that is very unlike do not find a value for *t'* for which $\text{Re}[\zeta((1+\varepsilon)/2 + it')]$ is lower than zero in the region of $x(t') \leq 0$ (note that $t_3$ is arbitrary).

## 3. Conclusions

In conclusion, the presence of zeros out of the critical line may lead to violation of the Heisenberg uncertain principle.

## References

[1] D. Schumayer and D. A. W. Hutchinson, Physics of the Riemann Hypothesis, Rev. *of Mod. Phys.* 83, (2011).

[2] I. Bakas and M. Bowick, Curiosities of arithmetic gases, *J. Math. Phys.* 32, 1881/1-4 DOI: 10.1063/1.529511 (1991).

[3] R. V. Ramos, F. V. Mendes, Riemannian Quantum Circuit, arXiv:1305.3759, (2013).

[4] H. Bohr, R. Courant, Neue Anwendungen der Theorie der diophantischen Approximationen auf die Riemannsche Zetafunktion, J. reine angew. Math. 144, 249-274 (1914).